\documentclass[pdflatex,aps,prl,superscriptaddress,twocolumn,floatfix]{revtex4}

\usepackage{graphicx}
\usepackage{dcolumn}
\usepackage{color}
\usepackage{wrapfig}

\begin{document}


\bibliographystyle{apalike}

\title{Reliability of regulatory networks and its evolution}


	\author{Stefan Braunewell}
	\author{Stefan Bornholdt}%
	\affiliation{%
	Institute for Theoretical Physics, University of Bremen, 
	D-28359 Bremen, Germany
	}%

\date{\today}


\begin{abstract}
The problem of reliability of the dynamics in biological regulatory networks
is studied in the framework of a generalized Boolean network model with 
continuous timing and noise. Using well-known artificial genetic networks such
as the repressilator, we discuss concepts of reliability of rhythmic attractors.
In a simple evolution process we investigate how overall network
structure affects the reliability of the dynamics. In the course of the evolution,
networks are selected for reliable dynamics.
We find that most networks can be easily
evolved towards reliable functioning while preserving the original
function.
\end{abstract}


\maketitle
\section{Introduction}
Biological systems are composed of molecular components and the interactions
between these components are of an intrinsically stochastic nature. At the same time,
living cells perform their tasks reliably, which leads to the question how
reliability of a regulatory system can be ensured despite the omnipresent
molecular fluctuations in its biochemical interactions.

Previously, this question has been investigated
mainly on the single gene or molecule species level. In particular, different
mechanisms of noise attenuation and control have been
explored, such as the relation of gene activity changes,
transcription and translation
efficiency, or gene redundancy \cite{Ozbudak:2002qy,Raser:2005rt,McAdams:1999}.
Apart from these mechanisms acting 
on the level of the individual biochemical reactions, also features of the circuitry
of the reaction networks can
be identified which aid robust functioning \cite{Barkai:1997vn,ASBL1999,vDMMO2000}.
A prime example of such a qualitative feature that 
leads to an increased stability of a gene's expression level 
despite fluctuations of the reactants is negative 
autoregulation \cite{Becskei:2000uq}.
At a higher level of organization, the specifics of the
linking patterns among groups of genes or proteins
can also contribute to the overall robustness.
In comparative computational studies of several different
organisms, it has been shown that among those
topologies that produce the desired functional behavior
only a small number also displays high robustness
against parameter variations. Indeed, the experimentally
observed networks rank high among these robust
topologies \cite{Kollmann:2005fj,Wagner:2005,Ma:2006fk}.

%



However, these models are based on the deterministic
dynamics of differential equations. Modeling of the intrinsic
noise associated with the various processes in the network requires
an inherently stochastic modeling framework, such as stochastic
differential equations or a Master equation approach \cite{TO2001,Kepler12012001,Ozbudak:2002qy,Rao:2002fk}.
These complex modeling schemes need a large number of
parameters such as binding constants and reaction rates and
can only be conducted for well-known systems or simple
engineered circuits.
For generic investigations of such systems, 
coarse-grained modeling schemes have been devised 
that focus on network features instead of the specifics
of the reactions involved \cite{Bornholdt:2005tc}. 

To incorporate the effects of molecular fluctuations 
into discrete models,
a commonly used approach is to allow random 
flips of the node states. 
Several biological networks have been investigated
in this framework and a robust functioning of the
core topologies has been identified \cite{AO2003,Li2004,DB:2007}.
However, for biological
systems, the perturbation by node state flips appears to be
a quite harsh form of noise: In real organisms, concentrations
and timings fluctuate, while the qualitative state
of a gene is often quite stable.
A more realistic form of fluctuations than macroscopic (state flip)
noise should allow for microscopic fluctuations. This can be implemented
in terms of fluctuating timing of switching events \cite{KB2004,Chaves:2005lr,Braunewell:2007lr}.
The principle idea is to
allow for fluctuations of event times and
test whether the dynamical behavior of a given network
stays ordered despite these fluctuations.

In this work we want to focus on the reliability criterion 
that has been used to show the robustness of the yeast
cell-cycle dynamics against timing perturbations
\cite{Braunewell:2007lr} and
investigate the interplay of topological structure
and dynamical robustness. 
Using small genetic circuits we explore the concept
of reliability and discuss design principles of reliable
networks.

However, biological networks have not been engineered 
with these principles in mind, but instead have emerged
from evolutionary procedures. We want to investigate
whether an evolutionary procedure can account
for reliability of network dynamics.
A number of studies has focused on the question of evolution
towards robustness \cite{Wagner:1997,BS2000,Ciliberti:2007lr,szejka-2007,Aldana:lr}.
However, the evolution of reliability against timing fluctuations
has not been investigated.
First indications that network architecture can be evolved
to display reliable dynamics despite fluctuating transmission
times has been obtained in a first study in \cite{BB2007}. Using a deterministic criterion 
for reliable functioning, introduced in \cite{KB2004-2}, it was found
that small networks can be rapidly evolved towards fully
reliable attractor landscapes. Also, if a given (unreliable) attractor is chosen
as the ``correct'' system behavior, it was shown that with a high
probability a simple network evolution is able to find a network
that reproduces this attractor reliably, i.e.\ in the presence of noise.

Here, we want to use a more biologically plausible
definition of timing noise to investigate whether
a network evolution procedure can generate robust
networks. We focus on the question whether
a predefined network behavior can be implemented in a
reliable way, just utilizing mutations of the network
structure. We use a simple dynamical rule 
to obtain the genes' activity states,
such that the dynamical behavior of the system is completely
determined by the wiring of the network.

\section{Model description}

\subsection{Boolean dynamics}
A widely accepted computational description of molecular biological systems
uses chemical master equations and simulation of
trajectories by explicit stochastic modeling, e.g. through the
Gillespie algorithm \cite{Gillespie1977}. However, this
method needs a large number of parameters to completely describe
the system dynamics. Thus, for gaining qualititative
insights into the dynamics of genetic regulatory system 
it has proven useful to apply strongly
coarse-grained models \cite{Bornholdt:2005tc}. 

Boolean networks, first introduced by Kauffman \cite{Ka1969}
have emerged as a successful tool for qualitative dynamical modeling and
have been successfully employed in models of
regulatory circuits in various organisms such as {\it D. melanogaster} \cite{AO2003},
{\it S. cerevisiae} \cite{Li2004}, {\it A. thaliana} 
\cite{CarlosEspinosa-Soto11012004}, and recently {\it S. pombe} \cite{DB:2007}.

In this class of dynamical models, 
genes, proteins and mRNA are
modeled as discrete switches which
assume one of only two possible states. Here,
the active state represents
a gene being transcribed or molecular concentrations (of mRNA or
proteins)
above a certain threshold level.
Thus, at this level, a regulatory network is modeled as a simple network
of switches.

Time is modeled in discrete steps and the state of all nodes is
updated at the same time depending only on the state
of all nodes at the previous time step according to the
wiring of the network and the given Boolean function.

When such a system is initialized
with some given set of node states, it will in general follow
a series of state changes until it reaches a configuration that has been
visited before (finite number of states). Because of the deterministic
nature of the dynamics, the system has then entered a limit cycle and
repeats the same sequence of states indefinitely (or keeps the same state
in the case of a fixed point attractor).

\subsection{Stochastic dynamics}
In the original Boolean model there are two assumptions that are clearly
non-biological and are thus often criticized: the discrete time which implies
total synchrony of all components; and the binary node states which
prohibit intermediate levels and gradual effects.

There have been various attempts at loosening these assumptions
while keeping the simplicity of the Boolean models. It is a clear
advantage of Boolean models that they operate on a finite state space.
The synchronous timing, however, does not hold a similar advantage
apart from computational simplicity. Models that overcome this
synchronous updating scheme have been suggested in a variety
of forms. In \cite{Chaves:2006} different asynchronous schemes
are used in the model of the fruit fly.
The simplest asynchronous model keeps the discrete notion
of time but lets events happen sequentially instead of simultaneously.
A continuous-time generalization of Boolean models that is
inspired by differential equation models has been
suggested in \cite{KB2004}. Here, the discreteness of the node states
is kept but the dynamics take place in a continuous time.
In \cite{KB2004-2,BB2007} the limit of infinitesimally small
disturbances from synchronous behavior is investigated.

This concept of allowing variations from the synchronous
behavior will also be used in this work.
The principle idea is to use 
a continuous time description
and identify the state of the nodes at certain times with the discrete
time steps of the synchronous description \cite{Glass:1975b}. 
Further, an internal continuous variable is introduced
for every node and the binary value of the node is obtained from 
this continuous variable using a threshold function. Now a differential equation
can be formulated for the continuous variable. 

\begin{figure}
\includegraphics[width=4cm]{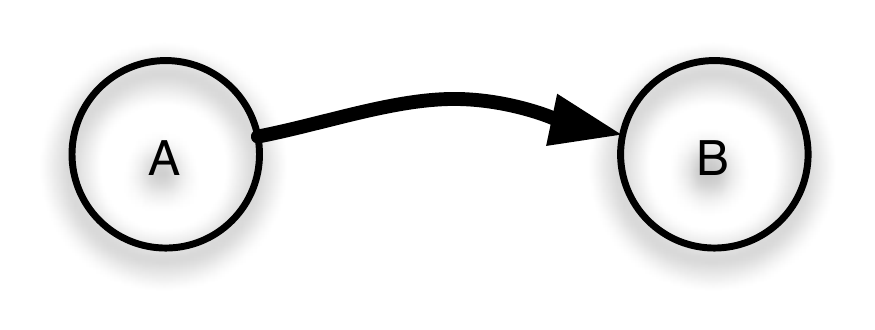}\\
\includegraphics[width=9cm]{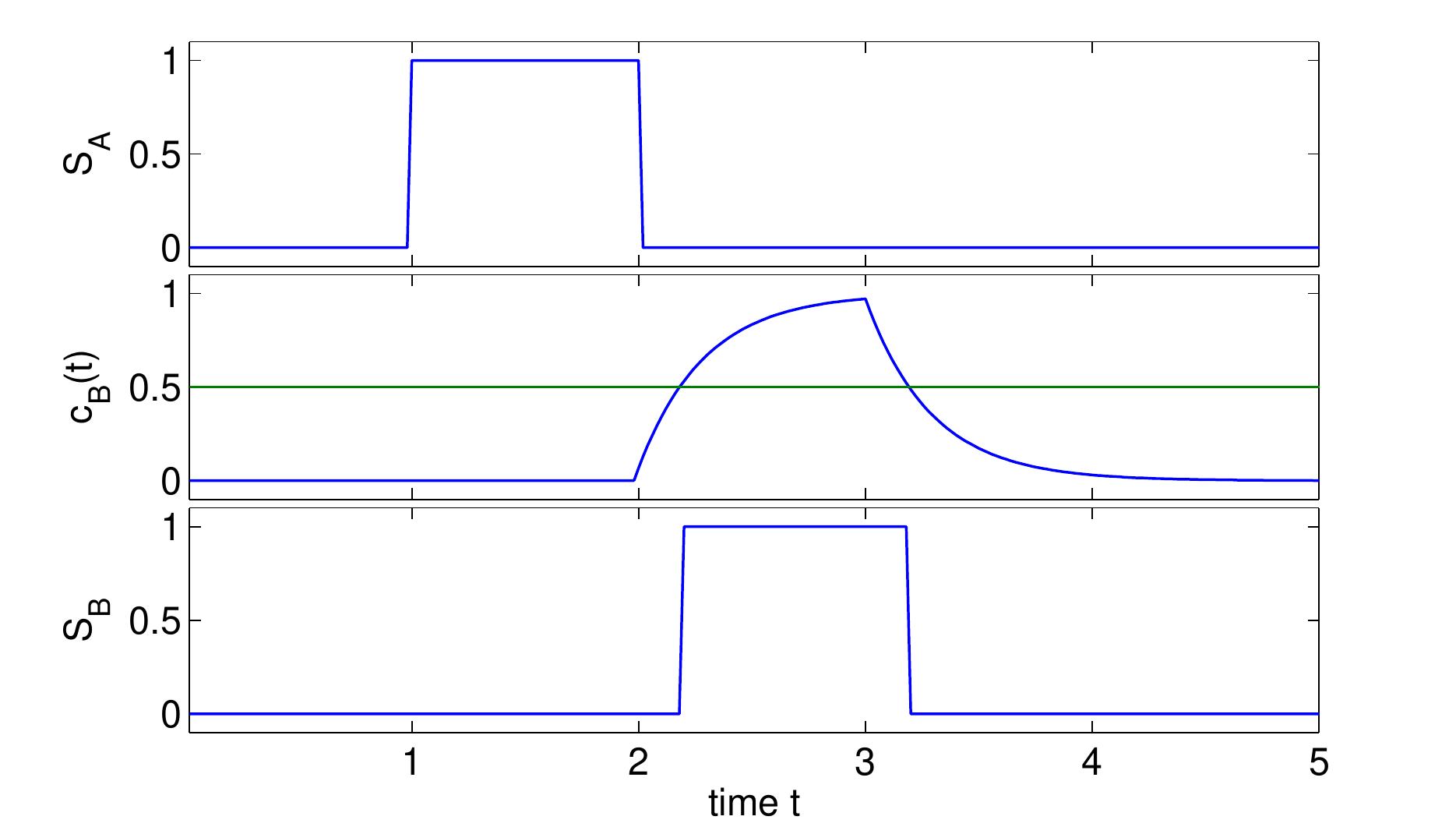}
\caption{Concentration buildup and decay of a protein given a specific 
input signal $S_A$ and the corresponding activity state $S_B$ ($t_d=1$, $\tau=0.3$).}
\label{fcharging}
\end{figure}

This is pictured in figure \ref{fcharging}. Here the internal
dynamics and the resulting activity state of a node with just one input are shown
for a given input pattern. The activator $A$ of the node $B$ is switched on (say, externally)
at time $t=1$
and stays on until it is switched off at time $t=2$. In the Boolean description we
would say node $A$ assumes state $S_A=1$ at time step 1 and at time step 2 switches to state $S_A=0$.
Node $B$ would react by switching to state $S_B=1$ at step 2 and to $S_B=0$ at step $3$.
In the continuous version, we implement this by a delay time and a ``charging'' behavior
of the concentration value of node $B$, driven by the input variable $S_A$. As soon as $c_B$ crosses the threshold of $1/2$,
the activity state of $B$ switches to $S_B=1$. 

Let us formulate the time evolution of a system of such model genes by the 
set of delay differential equations
\begin{equation}
\tau \frac{dc_i(t)}{dt}
=f_i(t,t_d)-c_i(t).
\label{eq:ODE}
\end{equation}
Here, $f_i(t,t_d)$ denotes the transmission function of node $i$ and describes
the effect of all inputs of node $i$ at the current time. The parameter $\tau$
sets the time scale of the production or decay process.
In general, any Boolean functions can be used as transmission function $f_i$. For simplicity, we choose threshold functions, which
have proven useful for the modeling of real regulatory networks \cite{Li2004,DB:2007}.

Let us use the following transmission function
\begin{equation}\label{updaterule}
f_i(t,t_d)=\left \{ 
\begin{array}{lll}
1,\quad &\sum_j a_{ij}S_j(t-t_d)\geq T_i,\\
0,\quad &\sum_j a_{ij}S_j(t-t_d)<T_i.
\end{array} \right.
\end{equation}
where $t_d$ is the transmission delay time that comprises 
the time taken by processes such as translation or diffusion 
that cause the concentration buildup of one protein to not 
immediately affect  other proteins. The interaction weight $a_{ij}$ 
determines the effect that protein $j$ has on protein $i$. An 
activating interaction is described by $a_{ij}=1$, inhibition 
by $a_{ij}=-1$. If the presence of protein $j$ does not affect 
expression of protein $i$, $a_{ij}=0$. The discrete state 
variable $S_i$ is determined by the continuous concentration variable $c_i$
via a Heaviside function $S_i(t)=\Theta[c(t)-1/2]$. The threshold value $T_i$ is given by 
$T_i=\sum_j a_{ij}/2$ (this choice is equivalent to the commonly used threshold value of 0
if the activity states are given by $S_i=\pm 1$ instead of the Boolean values used here).


For the simple transmission function given above, equation (\ref{eq:ODE})
can be easily solved piecewise (for every time span of constant 
transmission function), leading to the following buildup or
decay behavior of the concentration levels
\begin{equation}\label{solution}
c_i(t>t_0)=\left\{
\begin{array}{ll}
	1-(1-c(t_0))\exp(-(t-t_0)/\tau) & f_i \geq 0, \\
	c(t_0) \exp(-(t-t_0)/\tau) & f_i < 0.
\end{array}\right.
\end{equation}

This has the effect of a low-pass filter, 
i.e. a signal has to sustain for a while to affect the discrete 
activity state. A signal spike, on the other hand, will be filtered out. 

Up to now we have only introduced a continuous, but still deterministic generalization
of the synchronous Boolean model. If one now allows noise
on the timing delay, the model becomes stochastic and asynchronous.
The way we model this stochastic timing is via a signal mechanism. As soon as one
node flips its discrete state at, say, $t=t_0$, it sends a signal
to each node it regulates. This signal affects the input of a regulated node
at a later time $t=t_0+t_d+\chi$ where $\chi$ is a uniformly distributed random number
between $0$ and $\chi_\mathrm{max}$. The random number $\chi$ is chosen for each
signal and each link independently, which means that a switching node will affect two
regulated nodes at slightly different times.

Due to the timing perturbations, the network states at exactly
integer time do not hold a special significance any more.
To overcome this problem, we define a new macro step 
whenever all discrete node states 
(not the concentration levels) are constant
for at least a time span of $t_d/2+\tau$, which amounts
to one discrete time step in the synchronous model. Only
the system states at these times of extended rest
are used in the comparison with the synchronous behavior.

This way, small fluctuations of the signal events are tolerated, but extended times of
inactivity of the system must exist and the state of the network at these times must
correspond to the respective states under synchronous dynamics.
We call network dynamics ``reliable'' if, despite the stochastic effects
on the signal transmission times, the network follows the synchronous state
sequence. Although fluctuations in
the exact timing are omnipresent, ordered behavior of the sequence
of states can still be realized. An exact definition of the algorithm
can be found in the appendix.

In \cite{KB2004-2} a similar model, but with infinitesimal
timing perturbations, was used to identify those attractors
in Random Boolean Networks that are reliable. 
Most attractors in fact are unreliable, but are irrelevant for the system
because of very small basins of attraction. Further,
it was shown that the number
of reliable attractors scales sublinearly with
system size, which reconciles the behavior of Boolean dynamics
with the numbers of observed cell types as was originally
proposed in \cite{Kauffman:1993}. A similar result
was also obtained in a sequential updating scheme \cite{greil:048701}.

\section{Reliable and unreliable network dynamics}
%
%

\subsection{Dynamical sequence does not uniquely determine reliability}
\begin{figure}
\includegraphics[width=9cm]{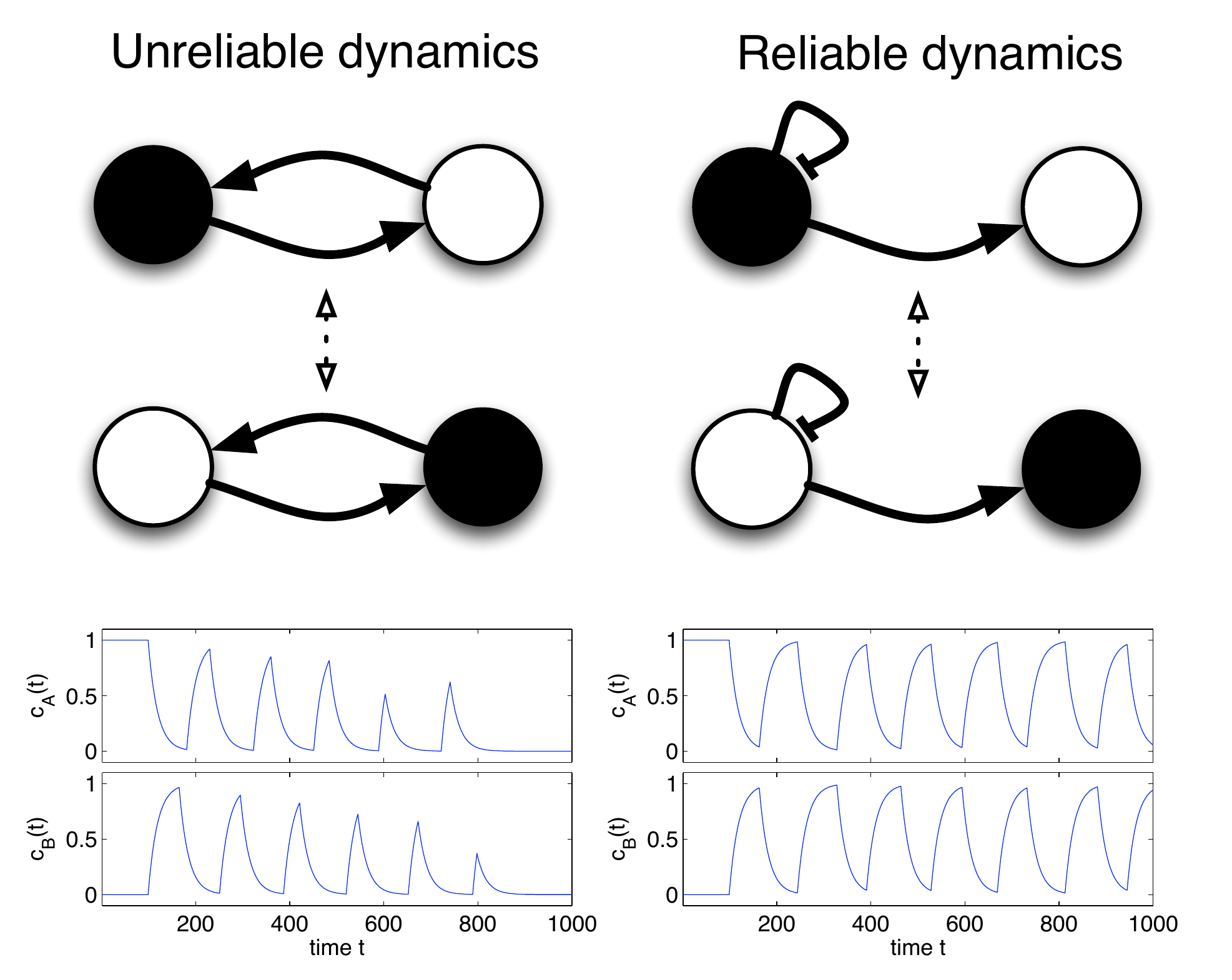}
\caption{Comparison of two networks that have a common synchronous attractor. The
mutually activating network is unreliable when subject to noise on the signal delay
times. In contrast, a negatively autoregulated gene that activates 
the second exhibits reliable dynamics.}
\label{reliableunreliablecomp}
\end{figure}

In this section we want to discuss the principle differences behind
reliable and unreliable dynamics and show that the same dynamical
sequence can be achieved
both in a reliable or in an unreliable fashion, depending on the underlying network
that drives the dynamics. \footnote{The parameters for the time delay, $t_d$, the production time constant,
$\tau$ and the noise level $\chi_\mathrm{max}$ are chosen for optimal
readability of the figures.
We stress that all our conclusions also hold for the parameter choice
from the results part or for any variation of these values within
reasonable bounds.}

We start with the well-known example of two mutually
activating genes and model the system according to equations
(\ref{eq:ODE}) -- (\ref{solution}). Apart from the trivial fixed points (both on or both off) this system
displays an unreliable attractor as shown in the left panel of figure \ref{reliableunreliablecomp}.
In the upper part, the synchronous Boolean attractor is depicted in a simple pictorial
form (black means active, white inactive). Below that the continuous variable
of both nodes is plotted over time in an example run and it can be seen that
because of desynchronization the system can exit the synchronous state sequence.

Changing just one link and thus creating an inhibiting self-interaction at the first gene
(see right panel of figure \ref{reliableunreliablecomp}), the dynamics
is now driven by this one node loop.
The synchronous sequence of the
attractor is still the same, but now the fixed points of the old network are 
no longer fixed points but transient states to this attractor.
The asynchronous dynamics, as shown in the lower part of figure \ref{reliableunreliablecomp}
now display an ordered behavior that would continue indefinitely.

The essential feature that causes these stable oscillations
is the time delay involved.
Without a time delay, the system would not exhibit stable oscillations
in either case but would assume intermediate levels
for both nodes. Thus, a direct comparison of these dynamics
with a stability analysis of ordinary differential equations without delay is not adequate.

\subsection{Biological examples}
\begin{figure}
\includegraphics[width=4cm]{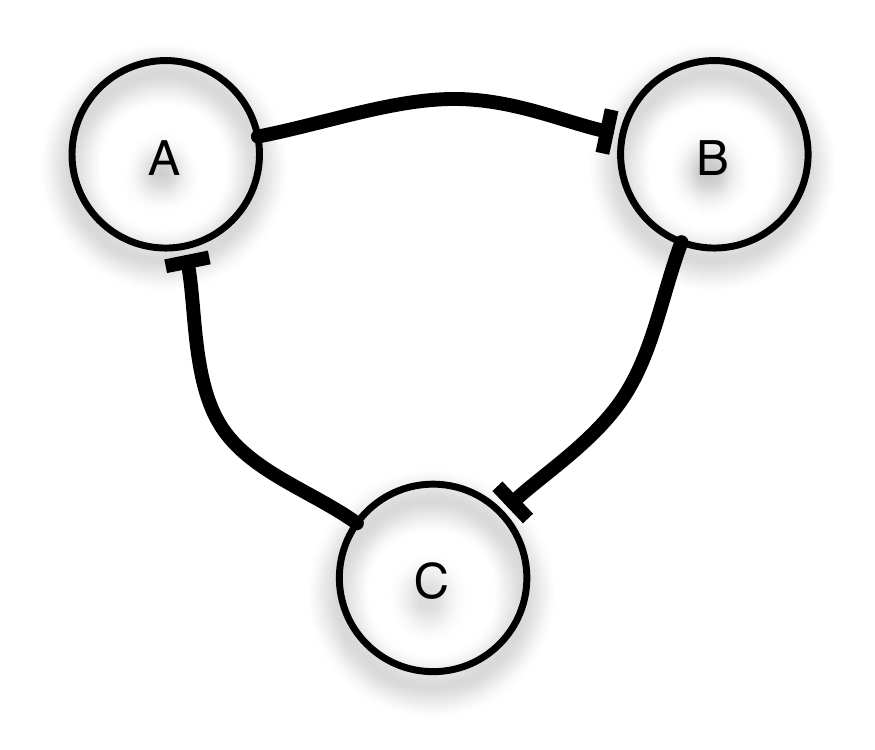}
\includegraphics[width=9cm]{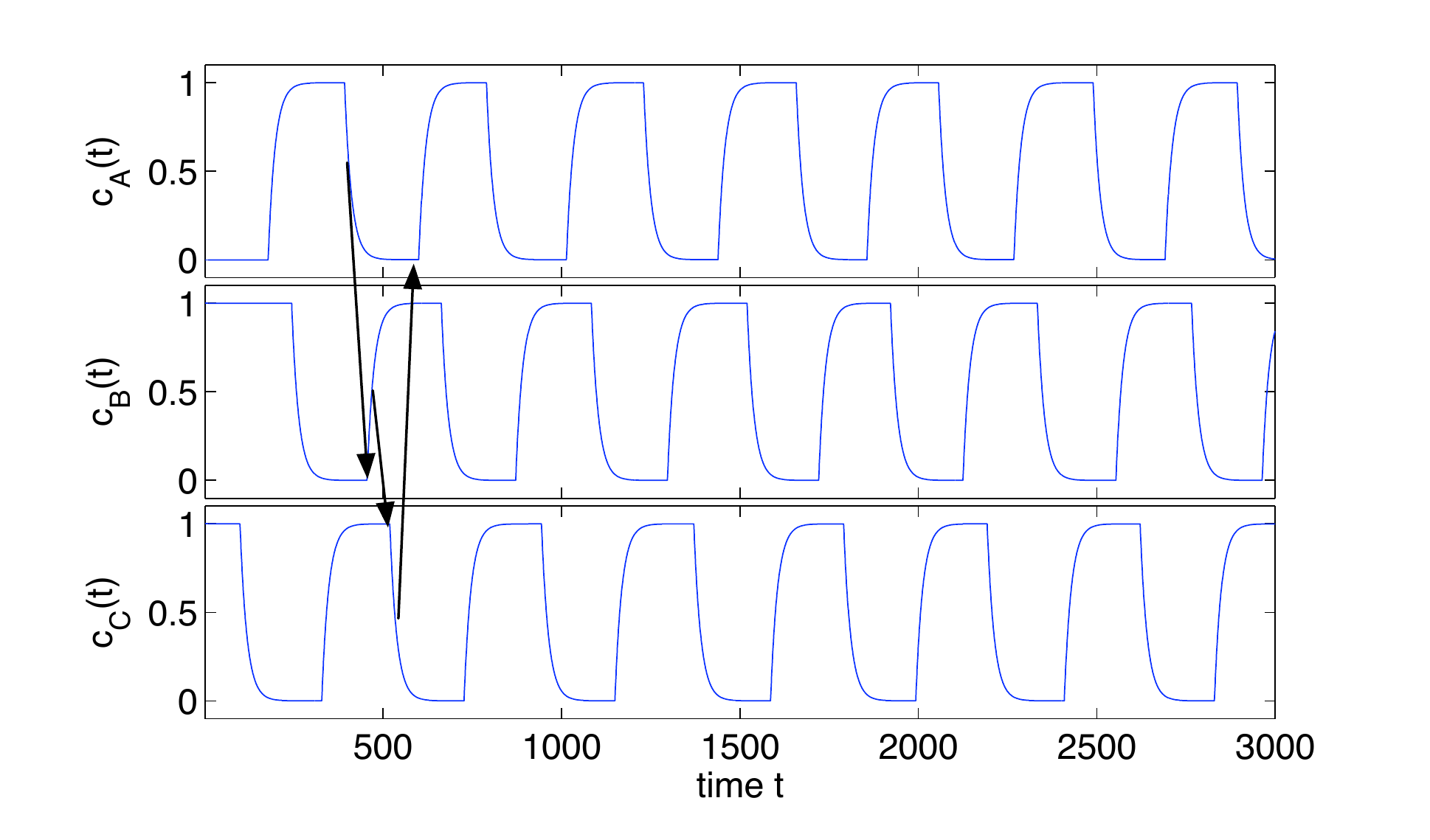}
\caption{Wiring diagram (top) and example time evolution of concentrations of all
three internal variables of the three-gene repressilator. The dynamics is governed by a single
event running around the circle -- here depicted by arrows which denote the flow of signals.}
\label{frep}
\end{figure}

Next, we now want to test the reliability of 
examples of circuits that can be created artificially. The repressilator
is a simple artificially generated genetic circuit implemented in
{\it E. coli} \cite{EL2000}. Consisting of 
of three genes inhibiting each other in a ring topology
(see upper part of figure \ref{frep}), this
system displays stable oscillations. 

Describing this system using differential equations
it was found that the unique steady state is unstable for
certain parameter values and that numerical integration
of the differential equations displays oscillatory behavior.
Also in a stochastic modeling scheme, sustained but irregular
oscillations can be observed, which show some
resemblance of the experimental time series \cite{EL2000}. 


To discuss this model system in our framework, the synchronous
Boolean description has to be analyzed first.
Here, the three-gene repressilator
exhibits two attractors which comprise all eight network states -- the ``all-active-all-inactive'' (two states) and
the ``signal-is-running-around'' pattern (six states). In the asynchronous scheme, independently
of the initial conditions, the system reaches the second attractor. Once the attractor is reached,
the system stays in it forever (i.e. is reliable in our definition) -- see figure
\ref{frep}. This is due to the fact that only a single switching happens at a given time. This
is depicted in the lower part of figure \ref{frep} by the arrows which are successively active,
no two events happening at the same time.

\begin{figure}
\includegraphics[width=4cm]{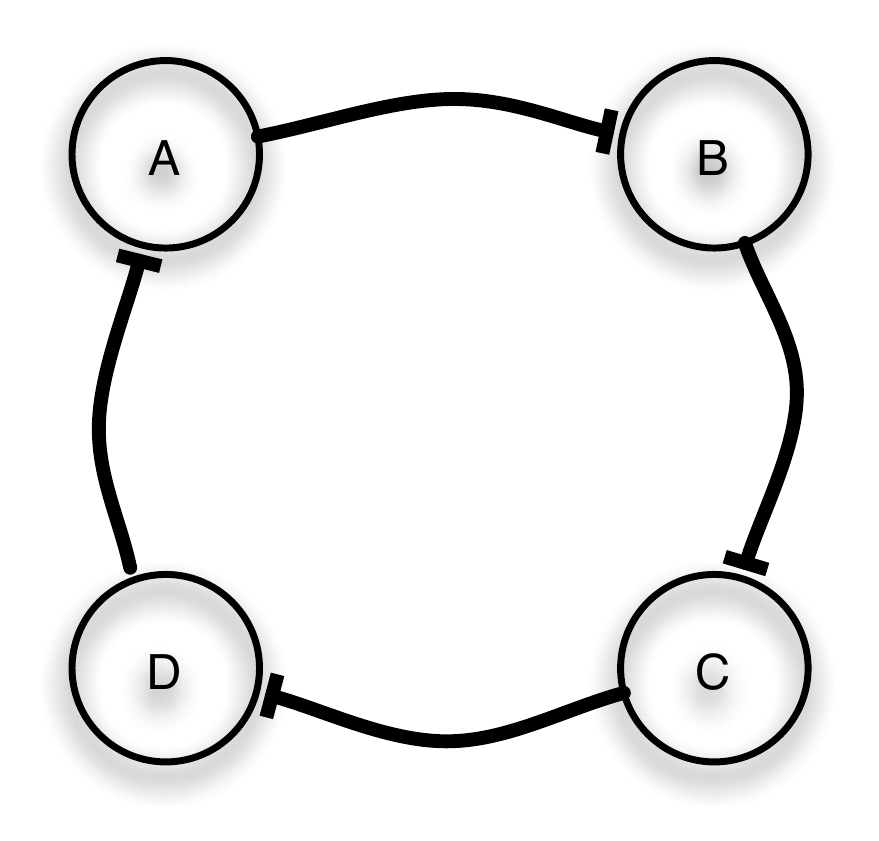}
\includegraphics[width=9cm]{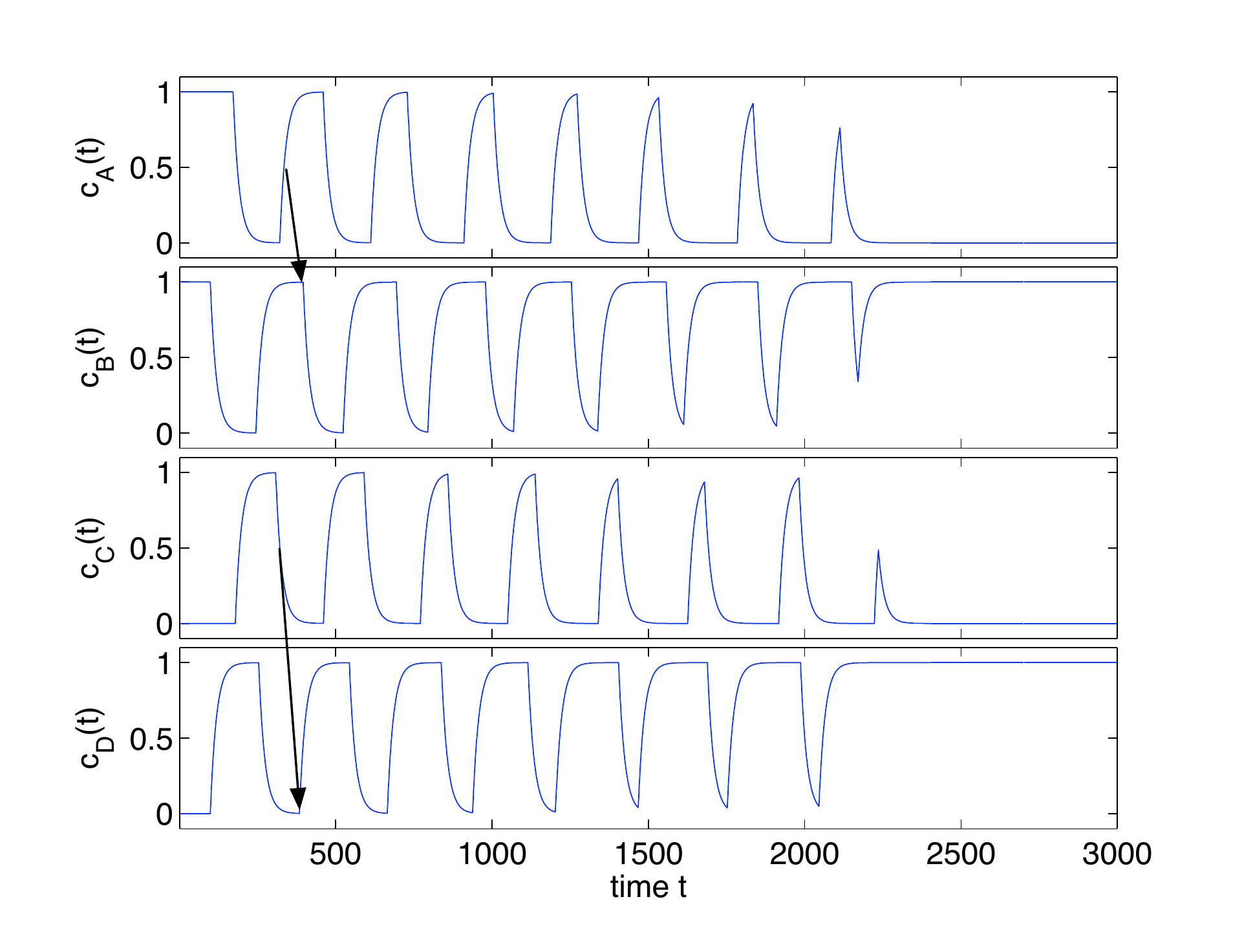}
\caption{Wiring diagram (top) and example time evolution of concentrations of all
four internal variables of the four-gene repressilator. As two events happen independently
at the same time (shown by the arrows depicting the signal events), 
the attractor can be left when the timing of the two event chains desynchronizes.
\label{f4rep}
}
\end{figure}

This picture changes in the case of the four-genes repressilator. The attractor structure
is now much more involved, consisting of the two fixed points (1010) and (0101) and three attractors
with four states each. Using the stochastic scheme as before, we find that only the fixed points
emerge as reliable attractors of the system. If any state of one 
of the 4-cycles is prepared as initial condition, the system thus 
always ends up in one of the two fixed points. In figure \ref{f4rep}, an
example run is shown which is initialized with the state (1100). 

Without any noise, the system would follow a four-state sequence 
consisting of all states where the two active nodes are adjacent and the
two inactive nodes are as well. However, if a small perturbation is allowed, the system can exit this attractor
as shown in the lower panel of \ref{f4rep}. Here, we have drawn 
two arrows showing two causal events happening at the
same time. In fact, there are two independent causal chains in the system dynamics. If these
two chains fluctuate in phase relative to each other, they can extinguish each other and drive the
system into a fixed point. For an extended discussion of the
concept of multiple causal chains, see \cite{KB2004}.

\subsection{Stable and unstable dynamics}
\begin{figure*}
\includegraphics[width=17cm]{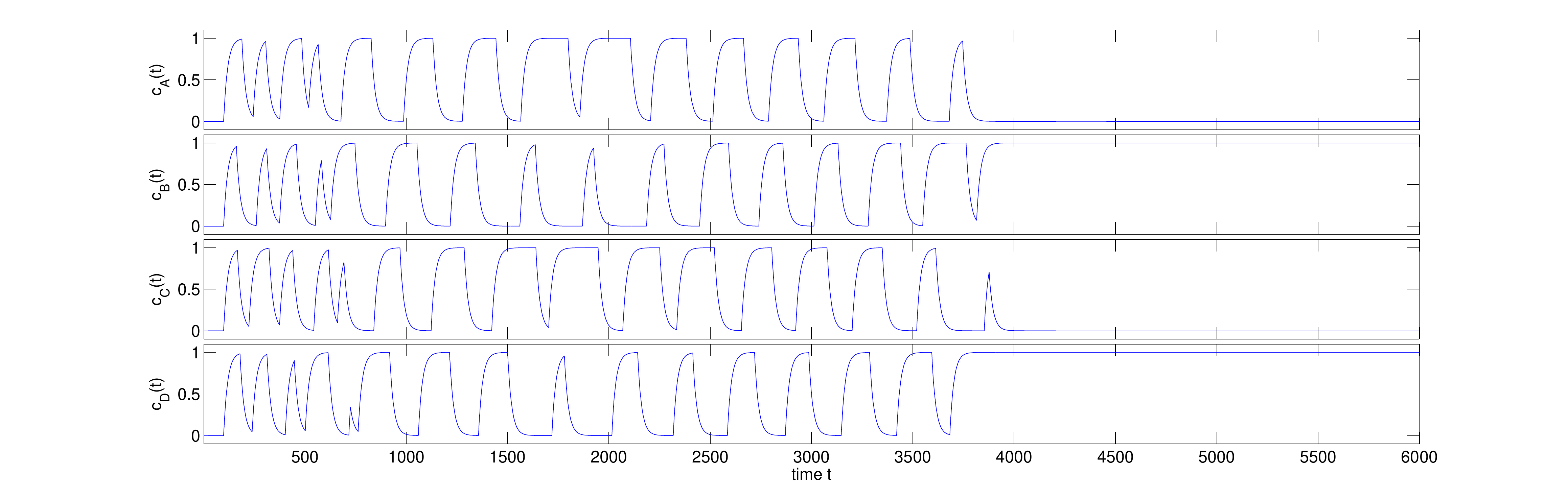}
\caption{Example of an unstable, a marginally stable and a fixed point
attractor in the four-gene repressilator.}
\label{frepunstable}
\end{figure*}

Apart from the concept explained in the last section, 
when a system can accumulate a phase lag through small perturbations
in a random walk-like fashion and eventually ends up in a different attractor,
there are also examples of systems, in which
any small perturbation drives the system away from the current attractor. 
This is the case if the concentration variables do not reach
their maximal value before their input nodes switch their state. This can
cause the following node to switch even earlier as compared to
the unperturbed timing (in the limit $\tau\to\infty$, this behavior is impossible and
each perturbation that is not removed from the system is merely ``neutral'').

We show this behavior in figure
\ref{frepunstable}. Here again, the four repressilator is shown, but with the initial state
configuration 0000. Without noise, this state belongs to the ``all-active-all-inactive'' attractor and
four independent events are happening at each time step. The small stochastic asynchrony
in the beginning is amplified and leads to a quick
loss of the attractor. The system then enters an intermediate attractor where the neutral
perturbation behavior is predominant, because the concentration levels have more time to
approach their saturation value.

The opposite behavior is also possible, that the system itself prevents divergence of the
phases. This can happen if the intermediate system state creates a signal spike
(i.e. a short-term status change of a node) that itself feeds back to the causal
chain. Even though the causal chains are independent in synchronous mode,
they can be connected through such intermediate states.

We want to stress that in the criterion employed in \cite{BB2007} 
it cannot be identified whether an attractor is marginally stable
or exhibits such a self-catching behavior. This is 
a limit of the deterministic criterion that is overcome by the explicit
modeling used here.

In this work, we consider all attractors as ``unreliable'' that can desynchronize
so strongly that the system does not maintain a ``rest phase'' in which no switching events
occur for an extended time. This includes all marginally
stable as well as unstable attractors. We do not distinguish between these in our results
as both do not seem suitable for the reliability of a biological system.






\section{Network evolution, simulation details}
Now that we have introduced the main concepts and ideas surrounding our definition
of reliability, we want to turn to the question, whether such a simple model of
regulatory networks can be evolved towards realizations displaying reliable dynamics. For this question,
we define the notion of a ``functional attractor''. As we are dealing with random networks,
we need a measure of what the system is supposed to do. Thus, we choose one
attractor of the starting network as the prototype dynamics that define the desirable dynamical
sequence. The functional attractor is determined by running the synchronous model with a randomly chosen initial state until an attractor is found.
During the evolution process we demand each network to reproduce
this attractor.

This prescription introduces a bias towards attractors with large basins of attraction.
However, as the basin of attraction is commonly understood as a measure of the
significance of an attractor, this appears to be a natural choice. Only unreliable attractors
are used as functional attractors, as in the case of reliable attractors, 
the evolution goal would be achieved before the start of the evolution.

The evolution procedure is chosen as a simple version of a mutation and selection process. 
We start by creating a directed random network with the prescribed number
of links $M$ (self-links are allowed)  and
determine the functional attractor. 
During the evolution procedure, we
mutate the current network by a single rewiring of a link, that is, removal of one link and
simultaneous addition of a random link between two nodes that are not yet
connected. This actually amounts to two operations on the graph, but has the
advantage that the connectivity $\langle k\rangle=M/N$ is kept fixed.

The fitness of a given network is assessed by comparison of the asynchronous dynamics
with the synchronous functional attractor. The initial network state is set according to
one randomly chosen state of the synchronous attractor. The concentration levels
are initialized to the same value (either $0.0$ or $1.0$).
Now the system dynamics is explicitly simulated using the stochastic algorithm
introduced above (details given in the appendix).
%
%
%
We follow the dynamics for maximally $10^6$ macro time steps (i.e.\ rest phases).
The fitness score is then obtained by dividing the number of steps 
correctly following the synchronous behavior by this
maximal step number.

The algorithm used in \cite{BB2007} is an abstraction of the principle
that small fluctuations cannot on its own drive the system out of its attractor,
but only if successively adding up. Thus, in a systematic way, for any possible
retardation of signal events it is checked whether it persists in the system
for a full progression around the attractor. If so, the attractor is marginally
stable and can in principle lose its synchrony. While this has the advantage
of being a deterministic criterion and thus leads to a noiseless fitness function,
there are situations in which this criterion is sufficient but not necessary
for reliable behavior.
By explicitly modeling the time course, as done in this work, only the truly unstable attractors
are marked as such.

We do not take into account the transient behavior of the system and define only
the limit cycle as the functional attractor. As our reliability definition would be trivially
fulfilled in the case of fixed points, we use networks exhibiting a limit cycle from the start.

In the evolution process, the fitness of a mutant is compared to the fitness that the
mother network scored. A network is only selected, if it scores higher than any
other network found before during the evolution. As the dynamics is inherently
stochastic, the fitness criterion is noisy, too. Thus, networks which are not
more reliable than the mother network might still be selected in the evolution
due to variability in the fitness score.

If a network follows the attractor up to a maximum
step number, it is said to be ``reliable''. If during the
network process a given number of mutation tries is exceeded, the evolution
process is aborted. 
The maximal number of mutation tries in the evolution is $a_\mathrm{step}=20000$
at each step
and $a_\mathrm{tot}=10^6$ during the full course of evolution. We later discuss the implications
of these fixed parameter settings.

We have used the following parameters in the results part. The delay time $t_d$ is set
to unity, the buildup time $\tau$ is $0.1$. Maximal noise $\chi_\mathrm{max}$ is $0.02$. 
This means, that the impact of any individual perturbation
is low and cannot itself cause a failure
in the fitness test. Only if several perturbations consecutively drive the system
away from synchronization, the requirement of an extended static period
can be missed.

\begin{figure}
\begin{center}
\includegraphics[width=8cm]{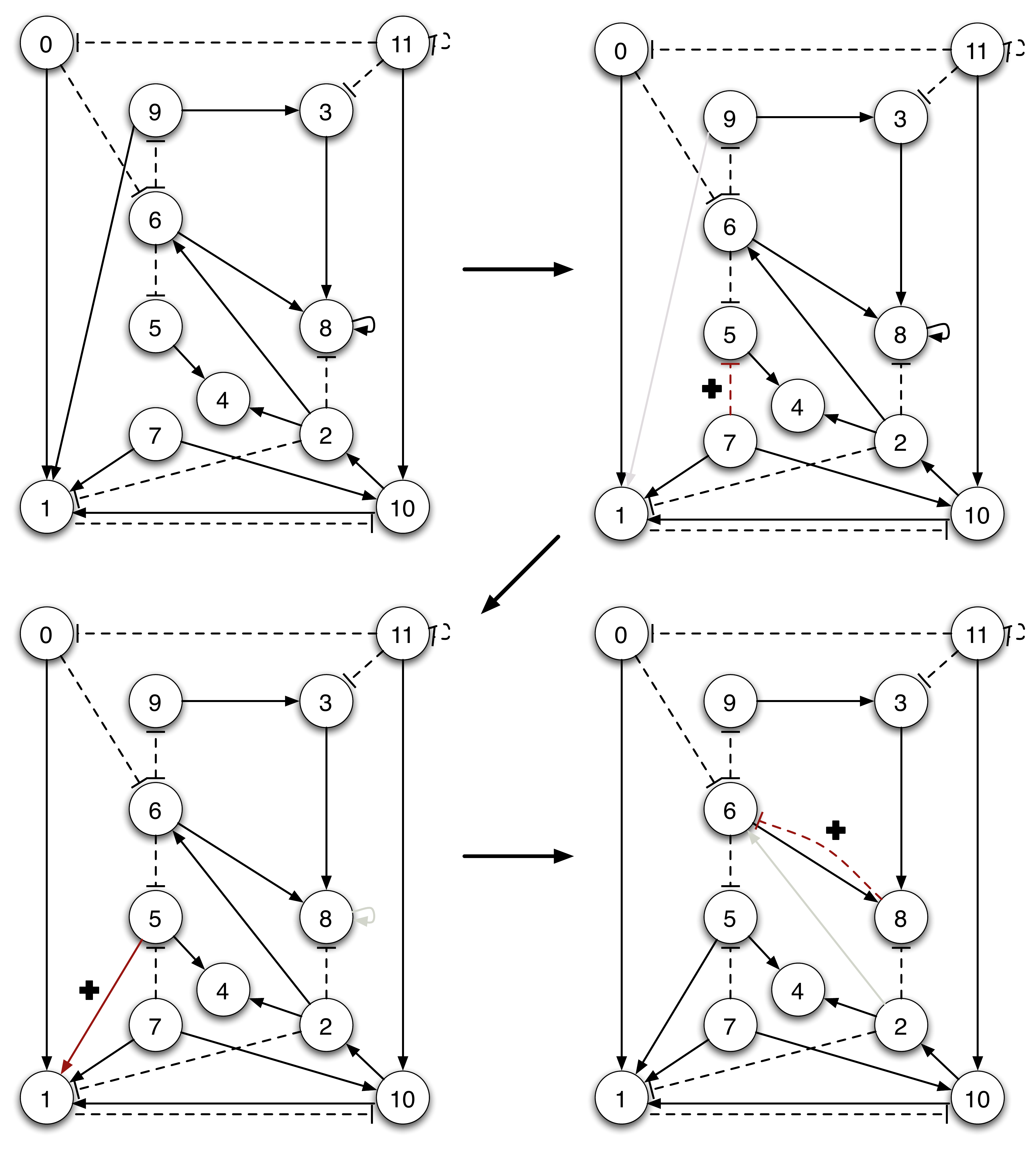}
\caption{A typical example of an evolution process for a network of size
$N=12$. In this example, three steps suffice for stabilization. 
The structure of each network during the evolution is shown, with the
arrows denoting the subsequent step in the evolution. In every step,
one link is lost (shown in grey color) and a new link is added (denoted 
by the plus sign). The change of the state space of the network
is given in figure \ref{fevoexampleattractors}.}
\label{fevoexamplestructure}
\end{center}
\end{figure}

\begin{figure}
\begin{center}
\includegraphics[width=8cm]{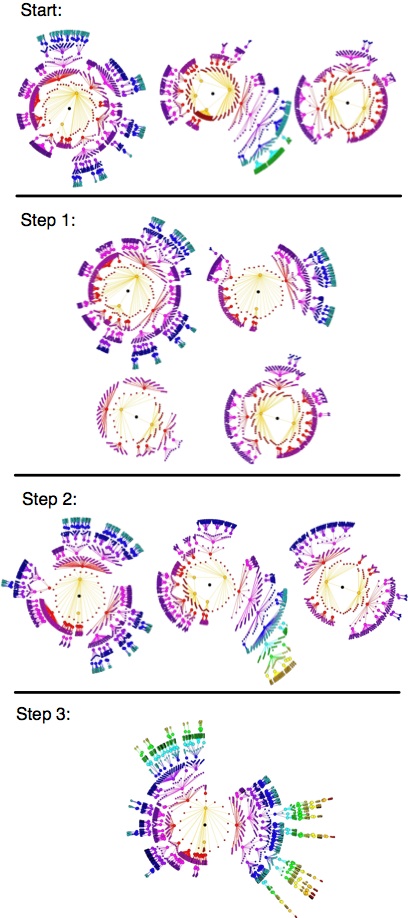}
\caption{Change of the (synchronous) attractor landscape during evolution --
corresponding to the network structures shown in figure \ref{fevoexamplestructure}.
For every step, the full attractor landscape is shown. Every dot denotes a state,
the subsequent state is connected via a line. The limit cycle is shown in the center
of each attractor basin.  The functional attractor is shown as the upper leftmost attractor in all steps.}
\label{fevoexampleattractors}
\end{center}
\end{figure}

In figures \ref{fevoexamplestructure} and \ref{fevoexampleattractors} we show an example of a typical
evolution process for a small network of $N=12$. During three steps, the network is evolved towards a reliable architecture. 
The initial network (upper-left in figure \ref{fevoexamplestructure}) displays three synchronous attractors (top panel in figure \ref{fevoexampleattractors})
of which the first is chosen as the functional attractor.
The structural changes are depicted in figure \ref{fevoexamplestructure} by a grey arrow for the removed
link and a plus-sign for the newly added link.
As is typical for these evolution processes \cite{BB2007}, the attractor landscape is affected dramatically
during the evolution. In this example, only the functional attractor survives the evolution procedure.

\section{Results of the network evolution}

\begin{figure}
\includegraphics[width=6cm,angle=-90]{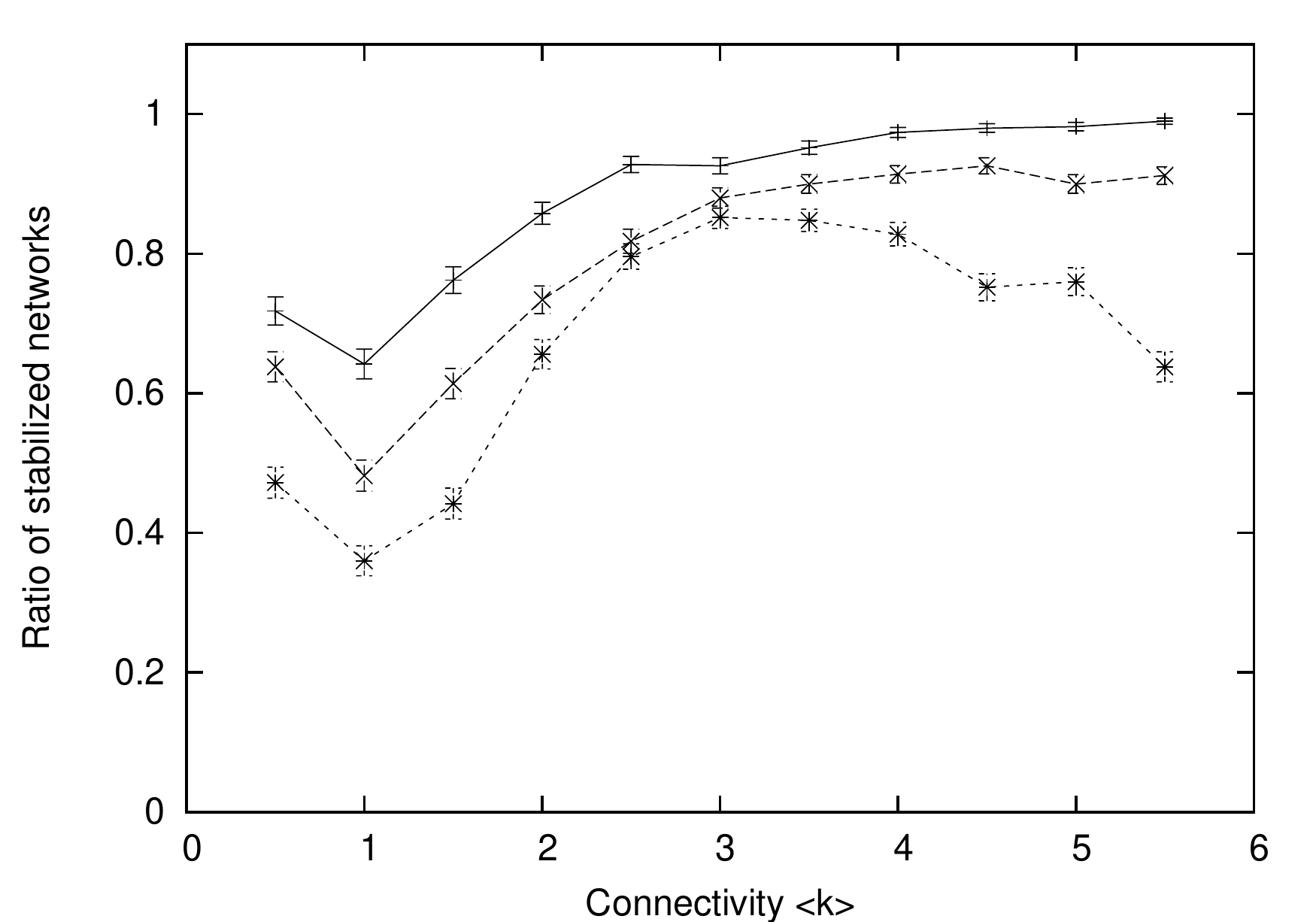}
\caption{Ratio of networks that were stabilized during the evolution
plotted against the average connectivity of the networks for network sizes
of N=16 (straigth line), N=32 (long dashes), N=50 (short dashes).}
\label{fevolution}
\end{figure}

We have performed the described network evolution for a variety of different
network sizes as well as connectivities. For system sizes of $N=16,32$ and $50$ and
connectivities between $0.5$ and $6$ the ratio of networks that 
were stabilized during the evolution is shown in figure \ref{fevolution}. Whenever
we plot the ratio of stabilized networks, we have calculated the sample errors by a Poissonian
error estimate, $\Delta x=\sqrt{x(1-x)/n}$, where $x$ is the obtained ratio from $n$ sample runs.

One can see that for intermediate connectivities between $2.5$ and $4.5$, 
the ratio of stabilized networks is above 80\% for all system sizes
under investigation.
This means, that starting from any random network, in four out of five cases
a simple network evolution is able to find a network that
displays the same dynamical attractor, but performs it reliably.

This result matches a very similar dependence on the connectivity
that was found for 16 nodes in the infinitesimal scheme in \cite{BB2007}.

It is interesting to note that for lower connectivities, the ratio of stabilized
networks decreases significantly. For all system sizes considered, there
is a sizable decrease of the stabilization ratio for connectivities below two.
This is especially apparent for the large system size $N=50$. This is due
to the ``essentiality'' of the structure on the dynamics. Changing a link
without destroying the
dynamical attractor is less likely for lower connectivities. 
At higher
connectivities the larger number of non-essential links
in the system aids evolvability towards reliable dynamics via phenotypically
neutral mutations.

However, considering large connectivities and large system sizes, 
the ratio of stabilized networks drops again. 
along with the increase of attractor lengths with system size that impairs
reproducibility of dynamics. Thus, we find an area of connectivity
between $1.5$ and $4.5$ for which the ratio of stabilized networks
is similar for all system sizes considered.

\begin{figure}
\includegraphics[width=6cm,angle=-90]{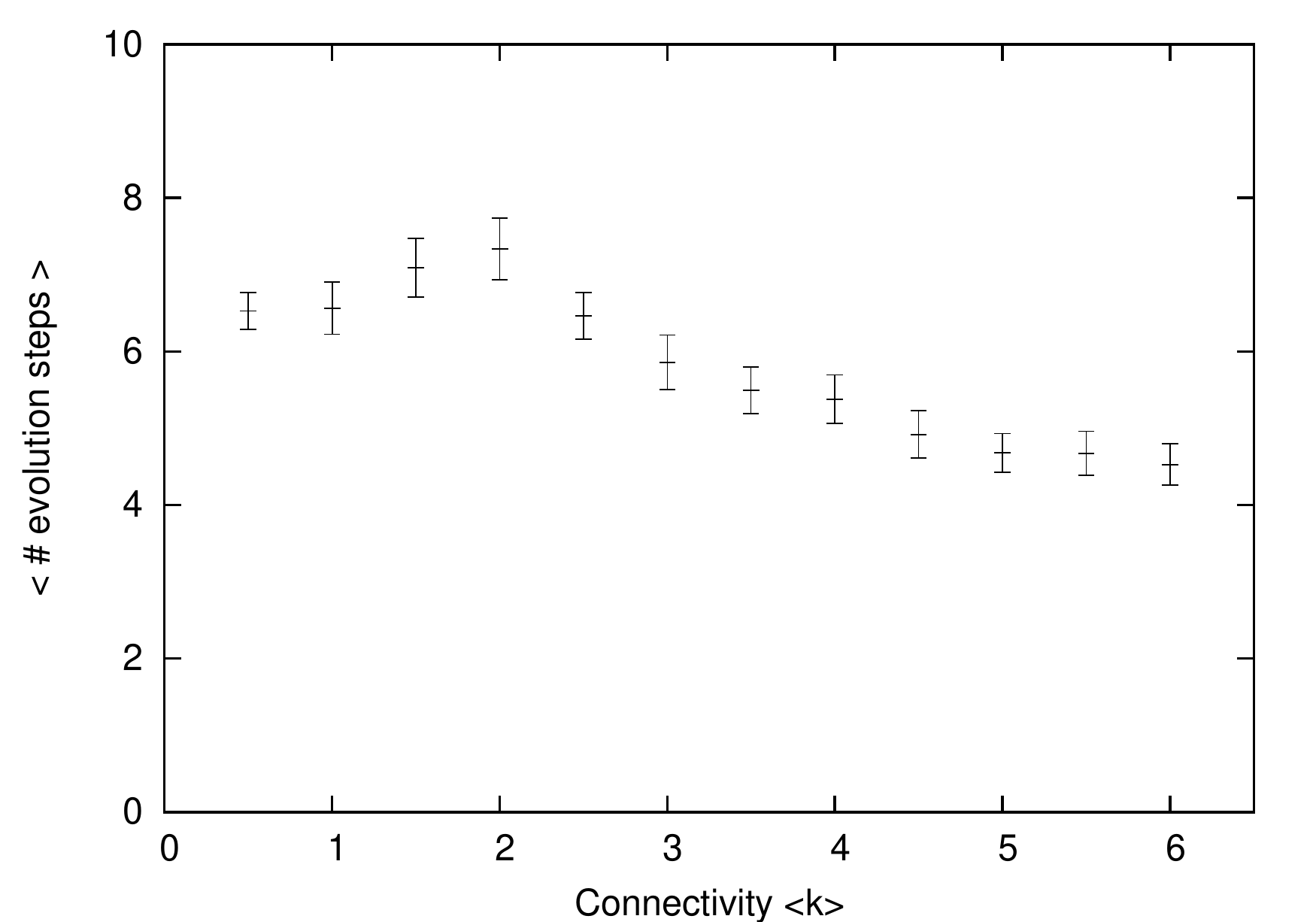}
\caption{Average number of evolution steps until stable realization is reached 
($N=32$)}
\label{fevo_steps}
\end{figure}

The plot in figure \ref{fevo_steps} shows the average number of rewiring
steps necessary until a stable network realization is found for networks
of 32 nodes. For all connectivities, this number is remarkably low,
as the evolution procedure basically implements a biased random
walk through structure space. This is due to the large variation
of the fitness score of a single network. Despite the rather small
evolutionary pressure, the evolution procedure quickly finds
a realization exhibiting reliable dynamics.
Interestingly, the number of evolution steps 
does not monotonically grow with the connectivity, but instead
drops for connectivities larger than two.

This again is an indication that networks with higher connectivities
are easier to evolve towards reliability. The ratio of 
links rewired in the evolution to the total number of links
is even monotonically decreasing (not shown).


\begin{figure}
\includegraphics[width=6cm,angle=-90]{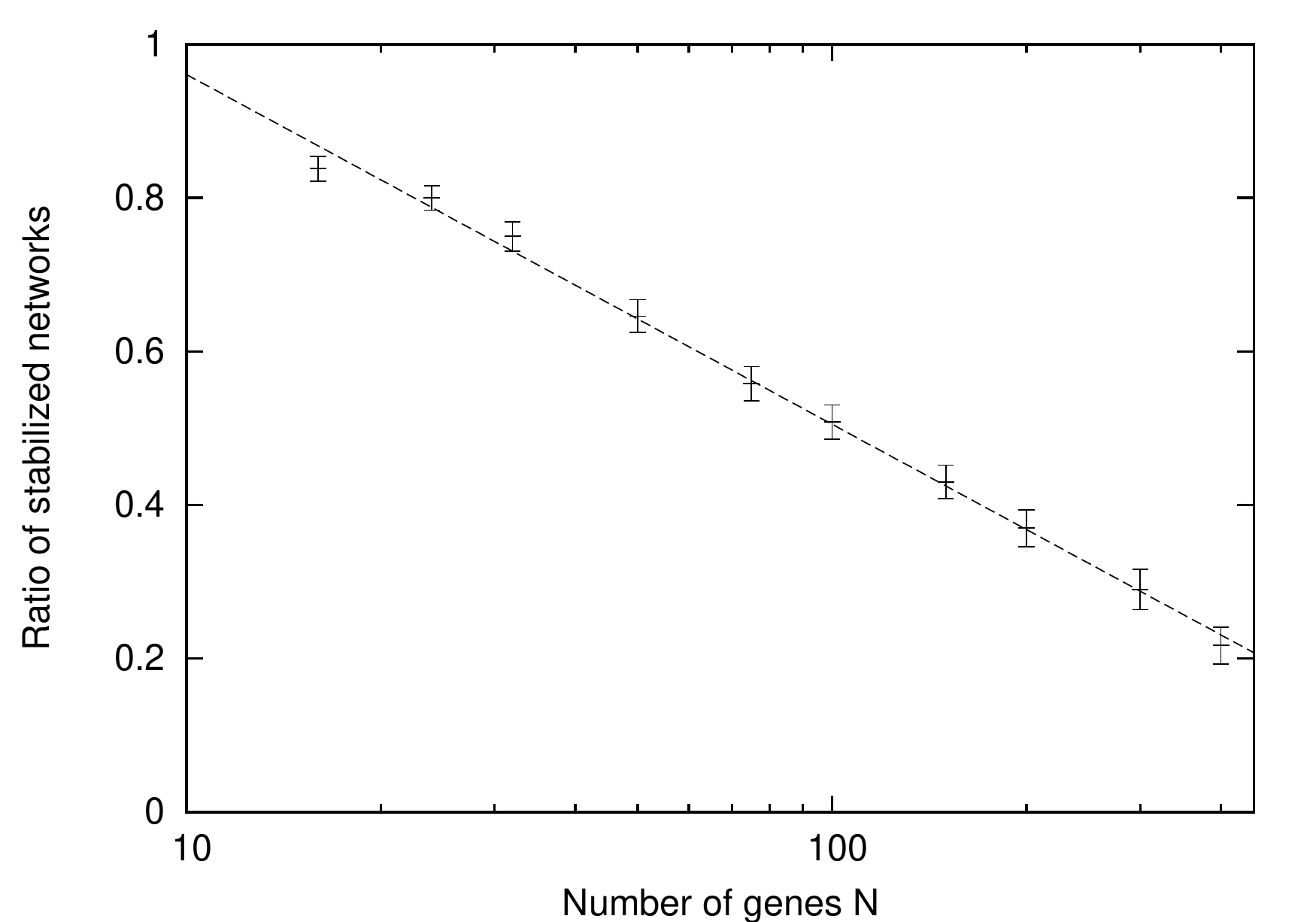}
\caption{Ratio of networks that were stabilized during the course of evolution
plotted against the number of nodes in the networks for an average connectivity of $\langle k\rangle=2$. 
The dashed line is given by a logarithmic fit of the data.}
\label{fevolutionNcomp}
\end{figure}

We want to further investigate the dependence
on network size by repeating
the evolution procedure with system sizes up to $N=400$.
This is shown in figure \ref{fevolutionNcomp} in a log-linear
plot of the ratio of stabilized networks vs.\ system size.
We find that the ability of the process to stabilize a given network
decreases with system size. The line in the figure represents
a fit of the function $f(N)=a-b\log(x)$ with $a=1.416$, $b=0.198$
thus a relatively slow decay with system size. One also has to keep
in mind that the fixed set of parameters for the number of
attempted mutations per evolution step and the total number of
attempted mutations during the evolution reduces the
success rate for larger networks. For small networks of $N=16$, 
20000 attempted mutations per evolution step suffices for a
good estimate of the space of all one-link mutations,
but as the number of possible mutations scales with the
system size N as $N^3$, it quickly becomes impossible to
check all possibilities. Thus, the results in figure \ref{fevolutionNcomp}
underestimate the probability to find a stable instance.


\begin{figure}
\includegraphics[width=6cm,angle=-90]{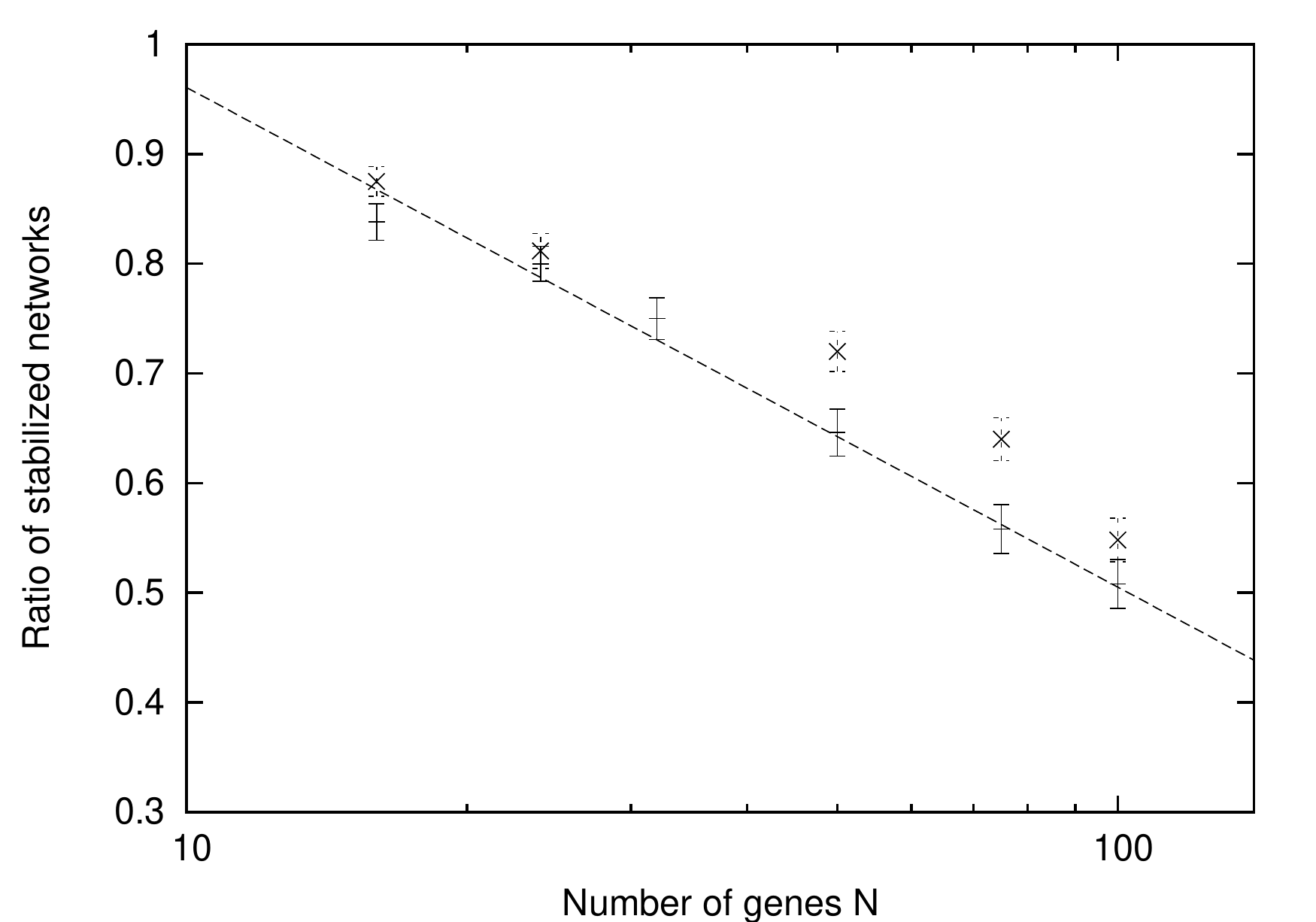}
\caption{Comparison of parameter values. 
Ratio of networks that were stabilized against the number of nodes in the networks for an average connectivity of $\langle k\rangle=2$. 
Original set of parameters marked with `$+$', points obtained with 
increased step number marked with `$\times$'.}
\label{fevolutionNcompparams}
\end{figure}

\begin{figure}
\includegraphics[width=6cm,angle=-90]{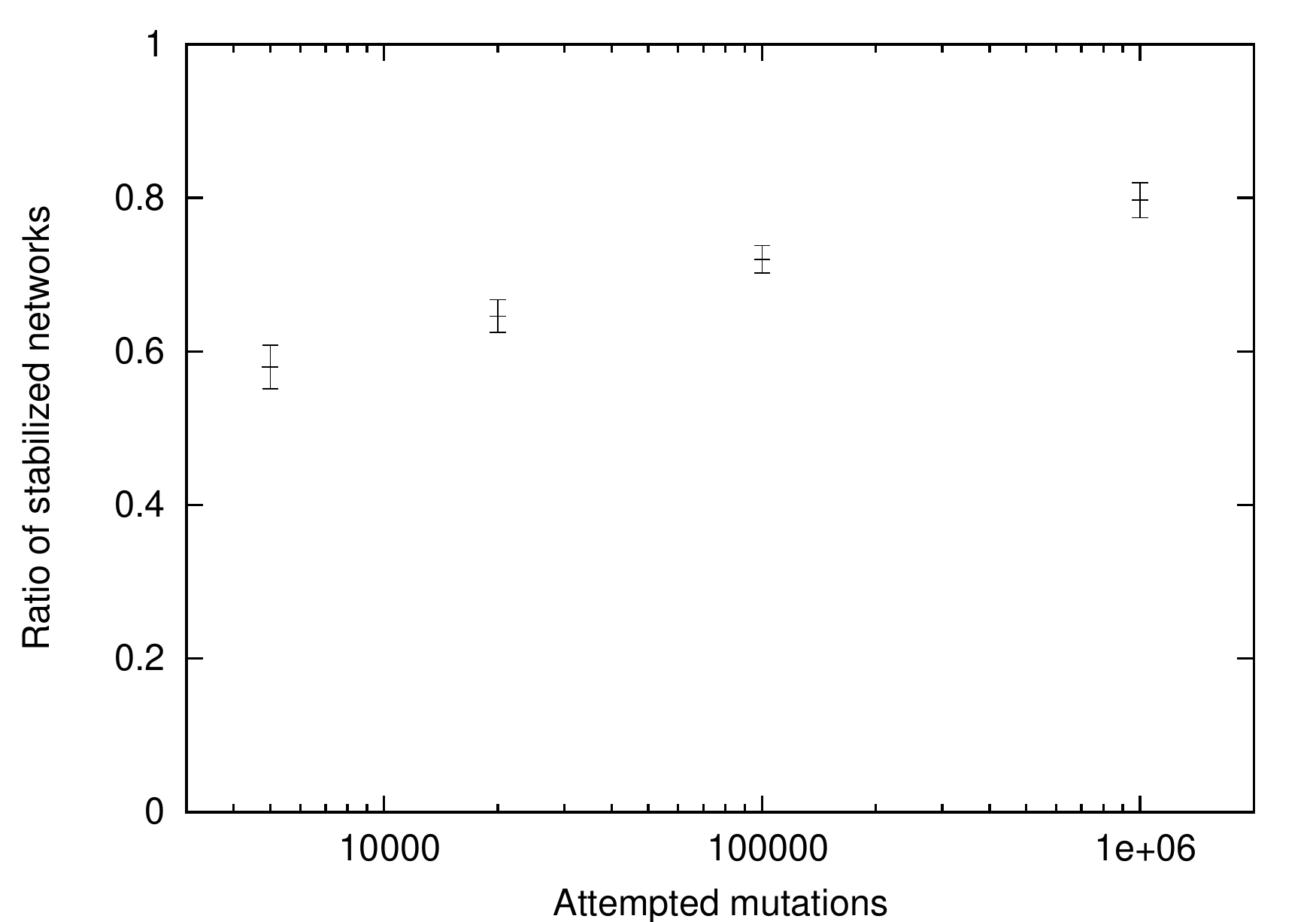}
\caption{Effect of parameter $a_\mathrm{step}$ on the results for $N=50$. The ratio of stabilized networks is plotted against the value of the paramter $a_\mathrm{step}$, giving the maximal number of attempted mutations per evolution step.}
\label{fevolution50Nparams}
\end{figure}

We have checked the dependence of the results 
on the selection parameters (attempted mutations per evolution
step $a_\mathrm{step}$ and total number of attempted mutations
during evolution $a_\mathrm{total}$)
for selected network sizes and connectivities. 
In figure \ref{fevolutionNcompparams} we again show the ratio of
stabilized networks against system size, but this time for two different
parameter values -- the original parameter set with $a_\mathrm{step}=2\cdot 10^4$
(denoted by `+') and for an increased value of $a_\mathrm{step}=10^5$ (denoted by `$\times$'). 
For small networks, the value of this parameter does not significantly
affect the results, but for $N>50$, differences can be clearly seen. For $N=50$ the
ratio rises from $0.65\pm0.02$ at $a_\mathrm{step}=2\cdot10^5$
to $0.72\pm0.02$ at $a_\mathrm{step}=10^6$. Interestingly, for larger system
sizes this effect does not seem to be amplified: for $N=100$ the
ratio rises from $0.51\pm0.02$ to
$0.55\pm0.02$. 

For $N=50$ we plot the dependence of the ratio of stabilized networks
on the parameter $a_\mathrm{step}$ in figure \ref{fevolution50Nparams}.
The largest parameter value used, $a_\mathrm{step}=10^6$ is about twice
the total number of possible rewirings and should thus suffice.

One can see that the decrease in the ratio of successfully
evolved networks can be significantly reduced when attempting
more mutations per evolution step. This is
due to the fact that an enormous number of mutations
is possible of which only a small fraction retains the
requested dynamical sequence.

Still, one can deduce from these results that it is harder to
stabilize large networks than smaller ones: even though there might
be a path to a stable network instance, it may not be practically
realizable as the chance to find exactly the right mutations may be too small. 

However, real world systems display
a large amount of modularity that leads to smaller cores of
strongly interacting components. We have not taken this into
account in our random network approach. We see this as a model
for small networks of key generators as were described in
recent Boolean models of biological systems \cite{Li2004,DB:2007}.
The resulting dynamics of the full networks are then influenced
by this core without strong feedback. This allows for rather
simple expression patterns of the full network without
constraints on the network size.

\section{Summary and Conclusions}
We have discussed a simple reliability criterion
for biological networks and have applied it
to network design
features that produce reliable dynamics.
We showed that small changes in the network
topology can dramatically affect the dynamical
behavior of a system and can lead to reliable
network dynamics.

To investigate how reliability can emerge 
in real-world systems that have been shaped by
evolution, we studied an evolutionary
algorithm that selects networks with a prescribed
dynamical behavior if they function more reliably
than a given mother network.

We found that a high ratio of random networks
can evolve towards instances displaying
reliable dynamics.
In accordance with other recent work
\cite{szejka-2007,Ciliberti:2007lr},
it was shown that the evolution of network structures can
lead to reliable dynamics both with a high probability
and within short evolutionary time scales.  

Surprisingly, small
connectivities are detrimental to this evolvability.
This is counter-intuitive as sparsely connected
networks show rather simple dynamics with short
attractor lengths. However, at the same time 
they are difficult to evolve because they have
a small structural ``buffer'' of links that can be
neutrally rewired without changing the dynamics.

This is related to the concepts of
``degeneracy'' and ``distributed robustness'' 
where additional elements are present in a system that are not
strictly necessary for the system's function but have a
positive effect on robustness \cite{GiulioTononi03161999,Wagner:2005b}.
Here, these additional elements are links that
are not strictly necessary to perform a specific
function. Thus, rewiring of these links
is possible and allows for a higher probability
to find a network with reliable dynamics.
We thus find in our framework that high
connectivity, although leading to
increasing complexity of the dynamics,
can be beneficial for the evolution of networks.

For larger system sizes the evolvability
towards reliable dynamics decreases.
This is due to the increasing
dynamical complexity of such networks
(longer attractor cycles, more non-frozen nodes).
Our strict criterion requests the
reliable reproduction of the exact state
sequence for every node, which leads to a
more difficult selection process
for large system sizes.


In summary, our results suggest that reliability
is an evolvable trait of regulatory
networks.
In the present simple model, reliability can be achieved by topological
changes alone and without fine-tuning of parameters.
This means that through mutations of the reaction networks,
biological systems may have the ability to rapidly acquire the property
of reliable functioning in the presence of biochemical stochasticity.

\section*{Acknowledgements}
The authors would like to thank Maria I.\ Davidich
for discussions and helpful comments on the manuscript
and Fabian Z\"ohrer for help with the state-space visualization graph.
This work was supported by Deutsche Forschungsgemeinschaft grants BO1242/5-1, BO1242/5-2.

\begin{appendix}
\section{Algorithm}
\small
The asynchronous algorithm is implemented such that
no discretized clock is needed. Only those times
will be investigated, when changes in the system happen.

For this, internal variables are needed to keep track of the dynamics. 
Every node $i$ has the following state variables:

\begin{itemize}
\item $t_{0,i}$: time of the last change of buildup/decay behavior
\item $c_i(t_{0,i})$: concentration level at that time
\item $b_i$: flag for current behavior - either buildup (1) or decay (0)
\item $s_{i,\mathrm{current}}$: current discrete state of node $i$
\item $s_{i,\mathrm{aim}}$: discrete state of node $i$ that would result
from the current states of all nodes: ${s_{i,\mathrm{aim}}=\Theta \left(\sum_{j=1}^n a_{ij} s_{j,\mathrm{current}} -1/2\right)}$,
\end{itemize}  

In addition, a global event queue $Q$ is maintained which keeps 
track of future changes in buildup/decay behavior.

The system is initialized by setting all values of discrete states 
$s_{i,\mathrm{current}}$ equal
to the state given by the discrete initial conditions. The concentration
levels are set to the same values ($0.0$ or $1.0$).
The times of the last behavior changes $t_{0,i}$ are set to $0$.

Before the simulation is started, for every node $i$ it is checked whether
the aspired state $s_{i,\mathrm{aim}}$ differs from the current state $s_{i,\mathrm{current}}$. 
If so, an event is added to the queue Q (sorted by time)
for time $t_d+\chi$, where $\chi$ is a uniformly distributed random
number between $0$ and $\chi_\mathrm{max}$. 

When the simulation is run, it is checked which of the two following
possible events takes place next:

\begin{enumerate}
\item Crossing of the concentration level of a node with the threshold
value $0.5$ 
\item The next event in queue $Q$
\end{enumerate}

A simple analytical expression can be given for the times when the concentration
levels are crossed (case 1). 
If $b_i=s_{i,\mathrm{current}}$, the node will not switch its state
because the concentration is moving away from the threshold. Otherwise,
one can calculate the time of the next concentration level to cross the
threshold by solving
equation (\ref{solution}) for $t$ with $c_i=0.5$:

\begin{equation}
\min_i \left[t_{0,i}+\tau \log(1+|1-2c_i(t_{0,i})|)\right]
\end{equation}

If an event of type 1 happens next, 
the discrete state of the respective node $i$, 
$s_{i,\mathrm{current}}$, is updated and the
effect on other nodes is calculated. For definiteness, let us assume this
crossing takes place at time $t$. If this switch causes the aspired
state of another node $j$ to switch, an event is sorted into the queue
$Q$ at $t+t_d+\chi$. When in the queue events for the
same node are scheduled to happen at later times, they will be removed.
They are thought to have been ``caught'' by the newly added event.

In the second case, the concentration level of the node
at time $t$ is
calculated according to equation (\ref{solution}) and saved as
$c_i(t_{i,0})$ with the new time $t_{i,0}$. 
The behavior flag $b_i$ is switched to reflect
that the node has changed from buildup to decay or vice versa.

If the time between any two successive node state changes 
in the network (not necessarily of the same node) is larger
than $t_d/2+\tau$, the node states are recorded and set as a new step
to be compared to the synchronous attractor.  
\end{appendix}

\end{document}